\documentclass[%
 floatfix,
 reprint,
 amsmath,amssymb,
 aps,
]{revtex4-1}
\usepackage{tipa}
\usepackage{siunitx}
\usepackage{graphicx}
\usepackage{dcolumn}
\usepackage{bm}
\usepackage{natbib}

\begin{document}


\title{Effective bandwidth approach for spectral splitting of solar spectrum using diffractive optical elements}

\author{Alim Yolalmaz$^{1, 2,}$}
 \email{alim.yolalmaz@metu.edu.tr}
\affiliation{
$^{1}$Programmable Photonics Group, Department of Physics, 
Middle East Technical University, 06800 Ankara, Turkey\\
$^{2}$Micro and Nanotechnology Program, Middle East Technical University, 06800 Ankara, Turkey\\
}%


\author{Emre Y\"{u}ce$^{1, 2, 3,}$}
\email{eyuce@metu.edu.tr}
\affiliation{
 $^{1}$Programmable Photonics Group, Department of Physics, Middle East Technical University, 06800 Ankara, Turkey\\
 $^{2}$Micro and Nanotechnology Program, Middle East Technical University, 06800 Ankara, Turkey\\
 $^{3}$The Center for Solar Energy Research and Applications (G\"{U}NAM), Middle East Technical University, 06800 Ankara, Turkey\\
}%


\date{\today}

\begin{abstract}
Spectral splitting of the sunlight using diffractive optical elements (DOEs) is an effective method to increase the efficiency of solar panels. Here, we design phase-only DOEs by using an iterative optimization algorithm to spectrally split and simultaneously concentrate solar spectrum. In our calculations, we take material dispersion into account as well as the normalized blackbody spectrum of the sunlight. The algorithm consists of the local search optimization and is strengthen with an outperforming logic operation called \textit{MEAN} optimization. Using the \textit{MEAN} optimization algorithm, we demonstrate spectral splitting of a dichromatic light source at 700 nm and 1100 nm with spectral splitting efficiencies of 92\% and 94\%, respectively. In this manuscript, we introduce an effective bandwidth approach, which reduces the computation time of DOEs from 89 days to 8 days, while preserving the spectral splitting efficiency. Using our effective bandwidth method we manage to spectrally split light into two separate bands between 400 nm - 700 nm and 701 nm - 1100 nm, with splitting efficiencies of 56\% and 63\%, respectively. Our outperforming and effective bandwidth design approach can be applied to DOE designs in color holography, spectroscopy, and imaging applications.

\end{abstract}

\maketitle


\section{\label{sec:level1}Introduction\protect 
\lowercase{} }
Spectral and spatial control of light has a strong influence in central fields such as imaging \cite{Mosk2012,Gjonaj2013}, sensing \cite{Konstantatos2010}, communications \cite{Wan2019}, and solar energy \cite{Ju2017, Mojiri2013} which directly affect the advancement of humanity. Traditional optical elements for controlling light such as lenses, mirrors, lamellar gratings, etc. are bulky for compact applications and expensive for large scale applications. However, modern approach of using spatially controlled pixels close to the wavelength of light drastically increases our ability to gain control on light. Spatiotemporal control of solar light yet is an important step towards increasing our benefits from solar radiation \cite{Xiao2016, Mojiri2013, Stanley2016}. 

The solar energy is an indispensable source of our era, and research in the effective conversion of the sunlight is vital for our future. Ground-breaking performance is especially obtained with solar cells that are fabricated via tandem architecture \cite{Green2013}. However, the tandem solar cells are costly to fabricate due to the complicated epitaxial growth procedure of multiple layers that impose constraints on design and performance. In addition to enhancing the performance of the solar cell materials and design architectures, concentration, tracking, and spectral splitting of the sunlight are the main routes to increase the efficiency of solar cells. Enabling sovereignty over light properties, spectral splitting is an important research field having functionalities in many applications \cite{Huang2016a, Basay2018}. The commonly used spectral splitting methods involve prisms, dichroic mirrors \cite{Mojiri2013}, holographic structures \cite{Riccobonoa}, combinations of prism-cylindrical lenses, and diffractive optical elements (DOEs) \cite{Huang2016a}. 

Being a versatile, low-weight, compact, easy-to-fabrication, and cost-effective, DOEs efficiently steer the incident light to a target position while simultaneously achieving spectral control of light. DOEs can effectively disperse the broadband light into its components via diffraction and multiple interferences. DOE, which is a result of computer-generated holography, could manipulate the light direction, wavefront, phase, and intensity \cite{Stanley2016}.

For designing a DOE there are a variety of algorithms: direct binary search \cite{Seldowitz1987}, iterative Fourier transform \cite{Kettunen2004}, Yang-Gu \cite{Yang1994}, Gerchberg-Saxton \cite{Gerchberg72}, and genetic optimization \cite{Johnson1993}. The design of broadband DOEs is more challenging due to the fact that they suffer from material dispersion that needs to be compensated. Besides, the number of design wavelength causes an expensive computational load. In addition to computational calculations, thanks to microfabrication technologies these structures can be fabricated at high precision and low cost \cite{Xiao2016, Lin2016, Huang2016}. As expected better precision in fabrication results in greater spectral splitting efficiency and beam concentration. However, fabrications of these structures with cheap photolithographic processes having less than \SI{1}{\micro\metre} resolution are not possible. Fortunately, with direct laser writing, such a resolution and large scale production is achievable \cite{Tokel2017}.

Here, we introduce a new optimization algorithm that outperforms the local search optimization algorithm. With accommodation of our algorithm, transmissive DOEs are designed for two purposes: solar concentrator and spectral splitter for the broadband light. In our model, we take material dispersion into account as well as the solar spectrum in order to provide a realistic structure. We manage to spectrally split and concentrate the solar energy into designated regions using a single DOE. Later, we introduce a bandwidth approach using which we optimize the DOE for every 13 nm wavelength steps within the 400 nm - 1100 nm bandwidth. We observe that optimizing the DOE for every 13 nm wavelength steps rather than a 1 nm wavelength step size within the target bandwidth results in similar splitting efficiency while the bandwidth approach provides 11 times faster computation.




\section{\label{sec:level2}Methods\protect 
\lowercase{} }
The local search optimization algorithm essentially adjusts the thickness of all DOE pixels in order to direct chosen frequencies to the target spots. However, the algorithm is sensitive to the choice of initial conditions. If the DOE thickness profile generated leads to a weak diffraction efficiency, the optimization algorithm may experience premature convergence or the optimization will last longer to converge to a satisfactory optimization state. Either to minimize or to eliminate these drawbacks, we propose the so-called \textit{OR}, \textit{AND}, and \textit{MEAN} logic operations as optimization criteria and always start with an initially random DOE structure. Considering the logic operations as \textit{OR}, \textit{AND}, and \textit{MEAN}, the algorithm updates the thickness profile of the DOE.

The local search optimization algorithm operates as follows: considering the size of the DOE, it generates a random 2D thickness profile. Then we calculate the spectral optical efficiency (SOE) and the enhancement for each design wavelength. We define the SOE as the ratio of intensity at the target area and the total intensity over the diffraction plane for a certain wavelength. Also, enhancement is defined as the ratio of SOE after and before the DOE. Next, the algorithm alters the thickness of each pixel from minimum pixel thickness to maximum one while monitoring increases in SOEs and enhancements for all design wavelengths. Each thickness scanning attempt of all pixels is called a sub-iteration, and all sub-iterations continue until the performance parameters are evaluated for each thickness of all pixels. The thickness profile of the DOE is updated in case a criterion which can be either \textit{OR}, \textit{AND}, and \textit{MEAN} is satisfied. As a decision making logic, \textit{AND} leads the algorithm to increase SOEs/enhancements of all design wavelengths during each sub-iteration. However, the algorithm with \textit{OR} criterion increases at least one of two design wavelengths during each sub-iteration. Lastly, the algorithm with \textit{MEAN} criterion updates DOE thickness profile if the mean of SOEs of all design wavelengths increases during a sub-iteration. After the contribution of all pixels is evaluated with scanning all thicknesses, we proceed to scan all pixels again until the enhancements/SOEs reach saturation providing no further increase in enhancement.

The DOEs in this study have 1600 pixels (40x40) with an individual pixel size of \SI{5}{\micro\metre}. A phase delay of an impinging coherent light at a wavelength of $\lambda$ on the output plane $\phi_{pq}$($\lambda$) is proportional to the thickness profile of the DOE $h_{pq}$, and is expressed as 
\begin{equation}
{\phi_{pq}(\lambda)=2\pi h_{pq}\lbrack n(\lambda)-1 \rbrack/\lambda},
\end{equation}
where $\lambda$ is the wavelength of the light source, n($\lambda$) the refractive index of the DOE, and $h_{pq}$ the thickness of the DOE at a position (p,q). The DOE profile is discretized to 8 levels with \SI{1}{\micro\metre} step size in thickness. Spectral splitting and beam concentration by a DOE depends on the step size and the spectral splitting efficiency increases with the number of step size as well as the number of pixels given the increased degrees of freedom to control. Here, the ultimate limit is set by the wavelength. Yet, fabrication of large-area DOEs at this precision is not feasible. In this study, we choose \SI{1}{\micro\metre} step size since this level of precision is easily achievable using direct laser writing \cite{Tokel2017}.

The output intensity profile of the DOE is computed by using discrete electric field summation presented in Eq. 2. This equation consists of the electric field of the incoming light on the target plane. In Eq. 2, $U_{pq}(\lambda)$ is the electric field at wavelength of $\lambda$ on the incident plane, $U_{ab}(\lambda)$ is the electric field at wavelength of $\lambda$ on the target plane, and $G_{pq}$ is the kernel transformation function at coordinates p and q. The electric field on the incident plane $U_{pq}(\lambda)$ and the transfer function $G_{pq}$ are defined in Eq. 3 and 4, respectively. In Eq. 3, $A_{pq}$ is the complex valued amplitude of the incident light wave at coordinates of p and q, $\phi_{pq}(\lambda)$ is the phase delay at wavelength of $\lambda$ after the DOE. In Eq. 4, (X,Y) and (x,y) represent the position of pixels at the target plane and the input plane, respectively and d is the distance between input and target planes.

\begin{equation}
{U_{ab}(\lambda)=\sum_{pq}U_{pq}(\lambda)*G_{pq}},
\end{equation}

\begin{equation}
{U_{pq}(\lambda)=A_{pq}(\lambda)*\text{exp}\left [ j\phi_{pq}(\lambda) \right ]},
\end{equation}

\begin{align*}
{G_{pq}=\sum_{pq}\left ( \frac{1}{j\lambda d} \right )*\text{exp}\left ( \frac{j2\pi d}{\lambda} \right )}
\end{align*}
\vspace{-4mm}
\begin{equation}
{*\text{exp}\left [j\pi\frac{\left \{  (y_{pq}-Y_{ab})^2+(x_{pq}-X_{ab})^2  \right \}}{\lambda d}  \right ]},
\end{equation}

In our calculations, we position the DOEs parallel to the output, and we do not take polarization, Fresnel reflection, absorption, and internal reflections within the DOEs into account. We include the dispersion curve of BK-7 for the DOE material \cite{BK7} and normalized the Blackbody radiation curve of the Sun at the surface temperature of 5778 K. The distance between the DOE and diffraction plane is chosen as \SI{350}{\micro\metre}.

\section{\label{sec:level3}Results and Discussions\protect 
\lowercase{} }
The performance of the aforementioned logic operations in the local search optimization algorithm is inspected by two light sources at first. The main goal here is to disperse the broadband spectrum into two regions between 400 nm - 700 nm and 701 nm - 1100 nm over target regions of size 0.1 mm - 0.2 mm, each. At first step, we spectrally split 700 nm and 1100 nm, which are the higher limits of the bands between 400 nm - 700 nm and 701 nm - 1100 nm, respectively. Given the fact that phase modulation of longer wavelengths is harder to manage (require thicker media for phase control), we choose to first control the longer wavelengths. We manage to spectrally split 700 nm and 1100 nm using a single DOE with a maximum thickness of \SI{8}{\micro\metre}. 

\begin{figure}
  \centering
  \begin{tabular}{cc}
    {\textbf{(a)}}\includegraphics[width=75mm]{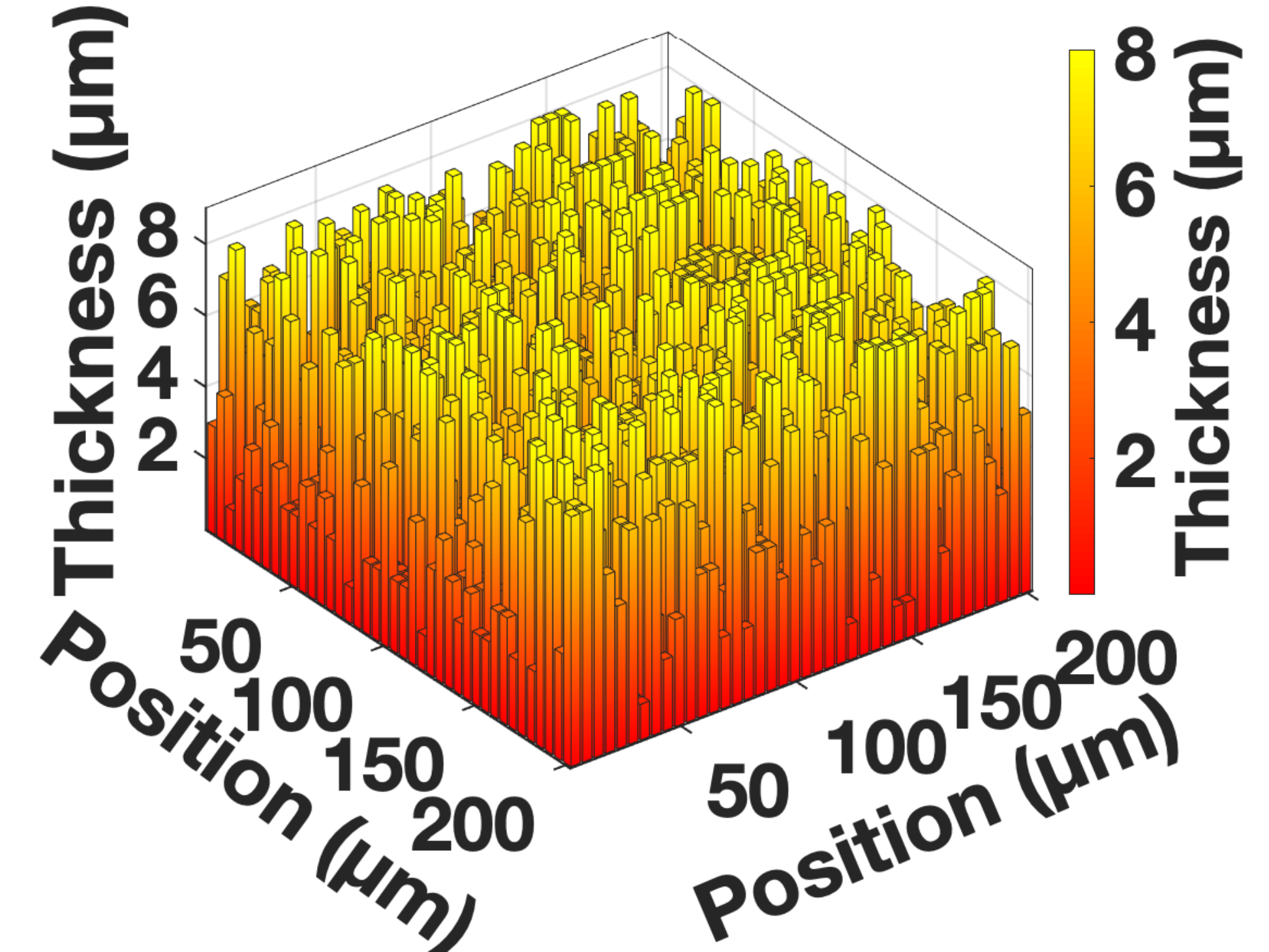}\\
    {\textbf{(b)}\includegraphics[width=75mm]{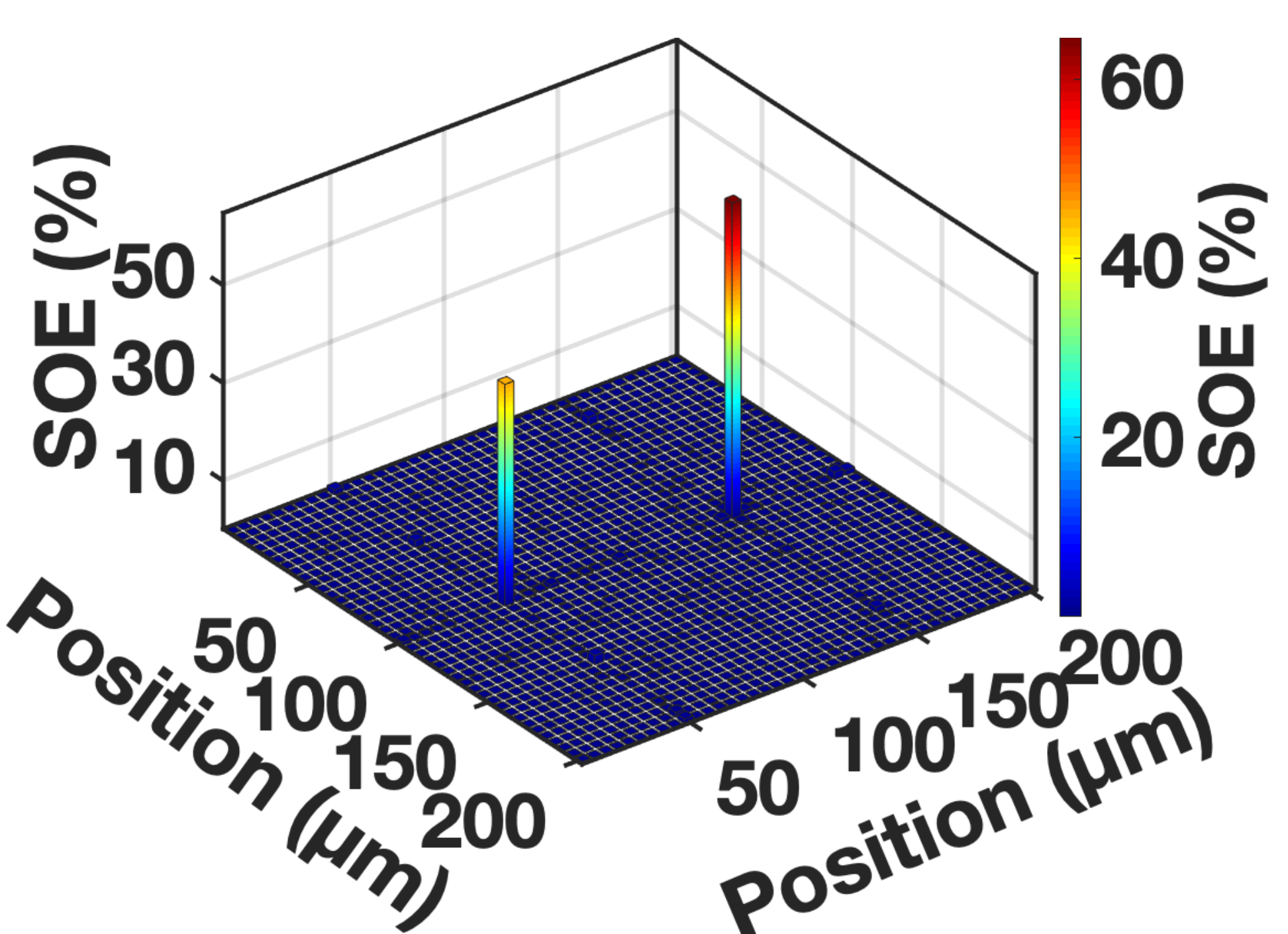}}\\
  \end{tabular}
  \caption{(a) Spatial distribution of DOE thickness for focusing two-color with the \textit{MEAN} criterion (the optimization criterion is set to rise the mean of SOEs of two sources), (b) Distribution of SOEs of incoming light on the target plane. The 700 nm light is guided to the position of (\SI{50}{\micro\metre}, \SI{100}{\micro\metre}) and of 1100 nm light is guided to the position of (\SI{150}{\micro\metre}, \SI{100} {\micro\metre}). The incident light sources at 700 nm and 1100 nm are successfully focused with 45\% and 63\% SOEs, respectively.}
  \label{distspots}
\end{figure}

In Fig. \ref{distspots}.a thickness profile of the DOE guiding two monochromatic waves of 700 nm and 1100 nm to targets as a result of \textit{MEAN} criterion is seen. Positions on the target plane and SOEs of these light sources are shown in Fig. \ref{distspots}.b. It is clearly seen that two wavelengths (at 700 nm and 1100 nm) are successfully directed to desired targets with 45\% and 63\% SOE, respectively. 

\begin{figure}
  \centering
  
  \begin{tabular}{cc}
    {\textbf{(a)}}\includegraphics[width=75mm]{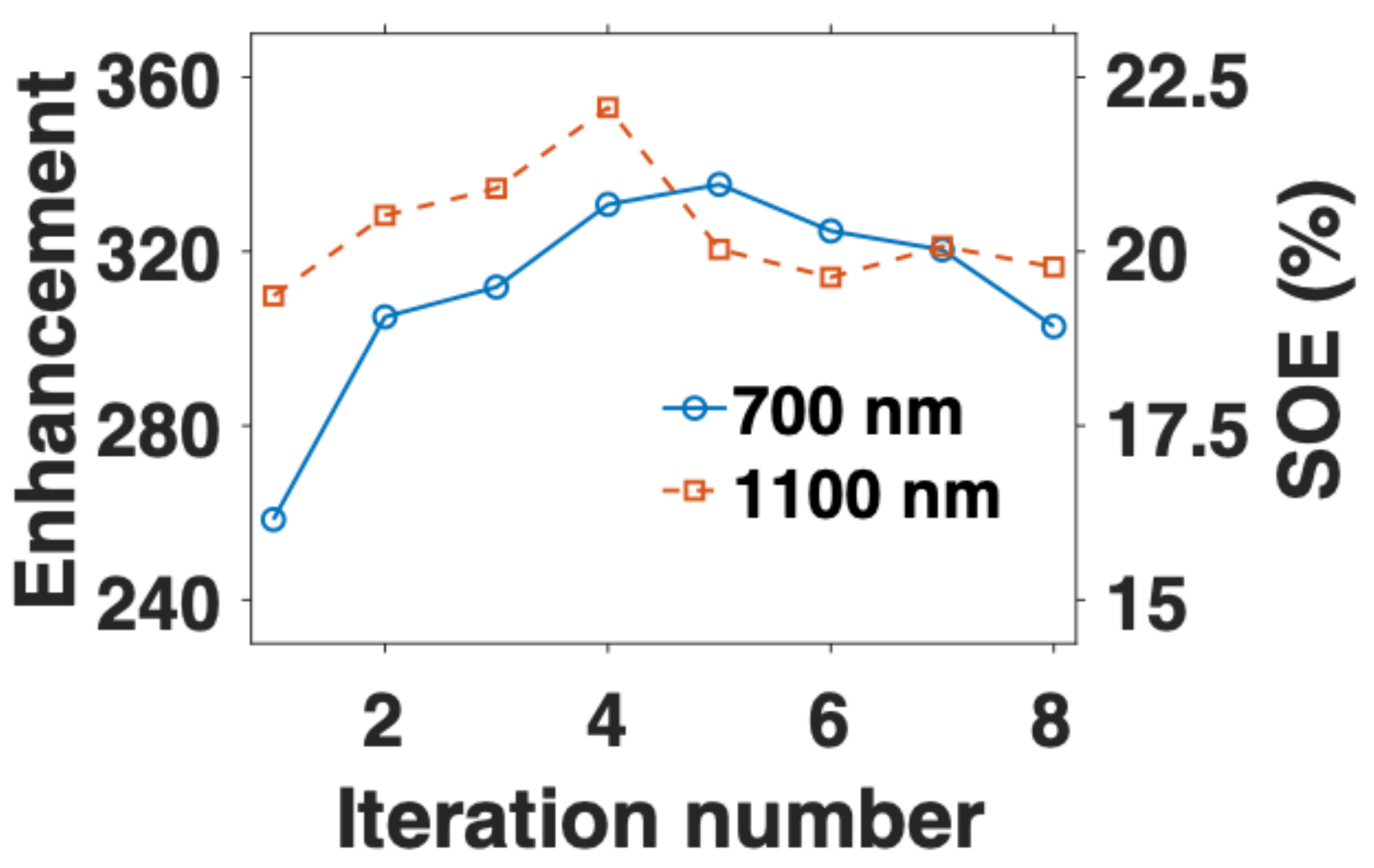}\\
    {\textbf{(b)}}\includegraphics[width=75mm]{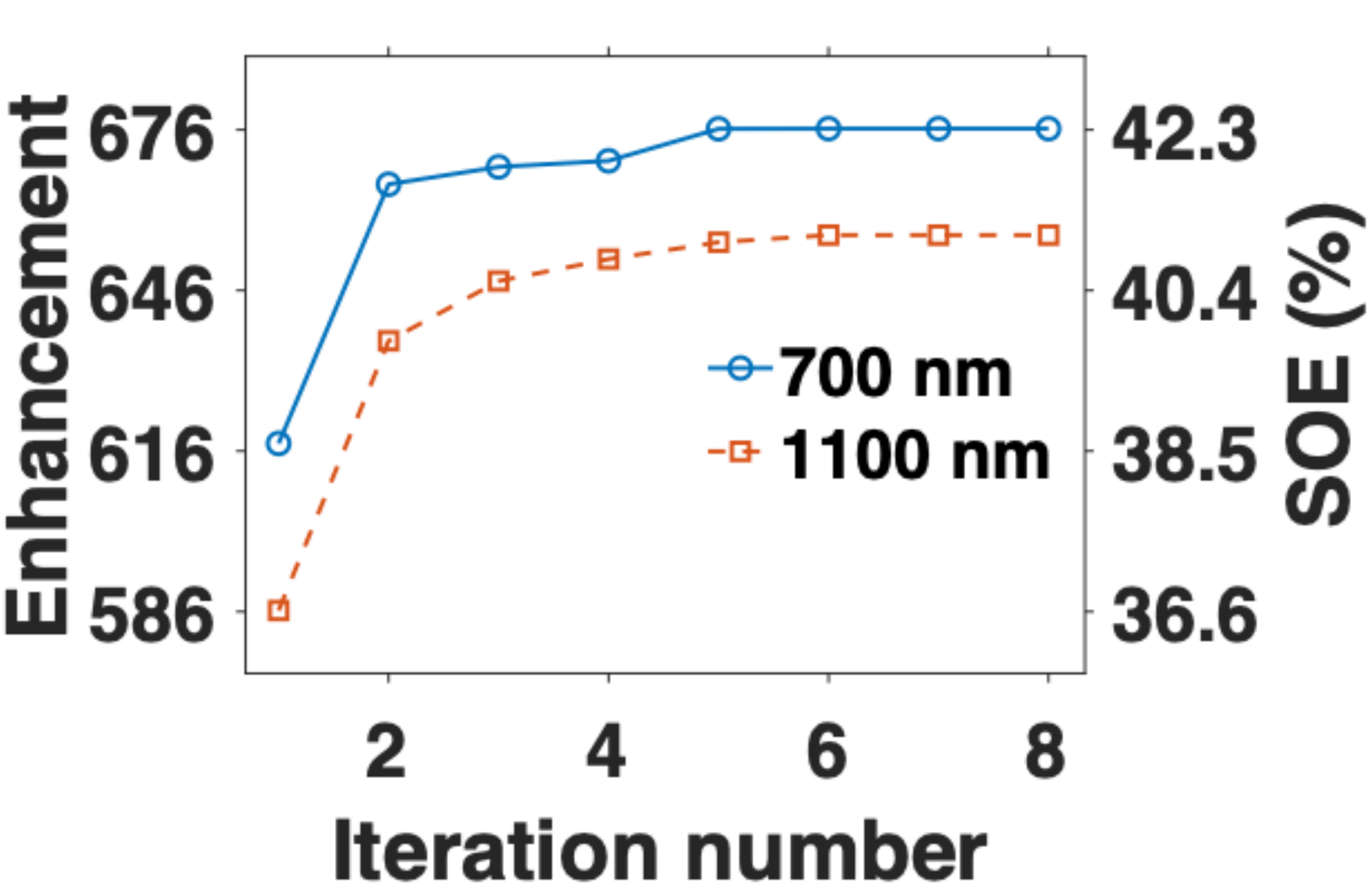}\\
    {\textbf{(c)}}\includegraphics[width=75mm]{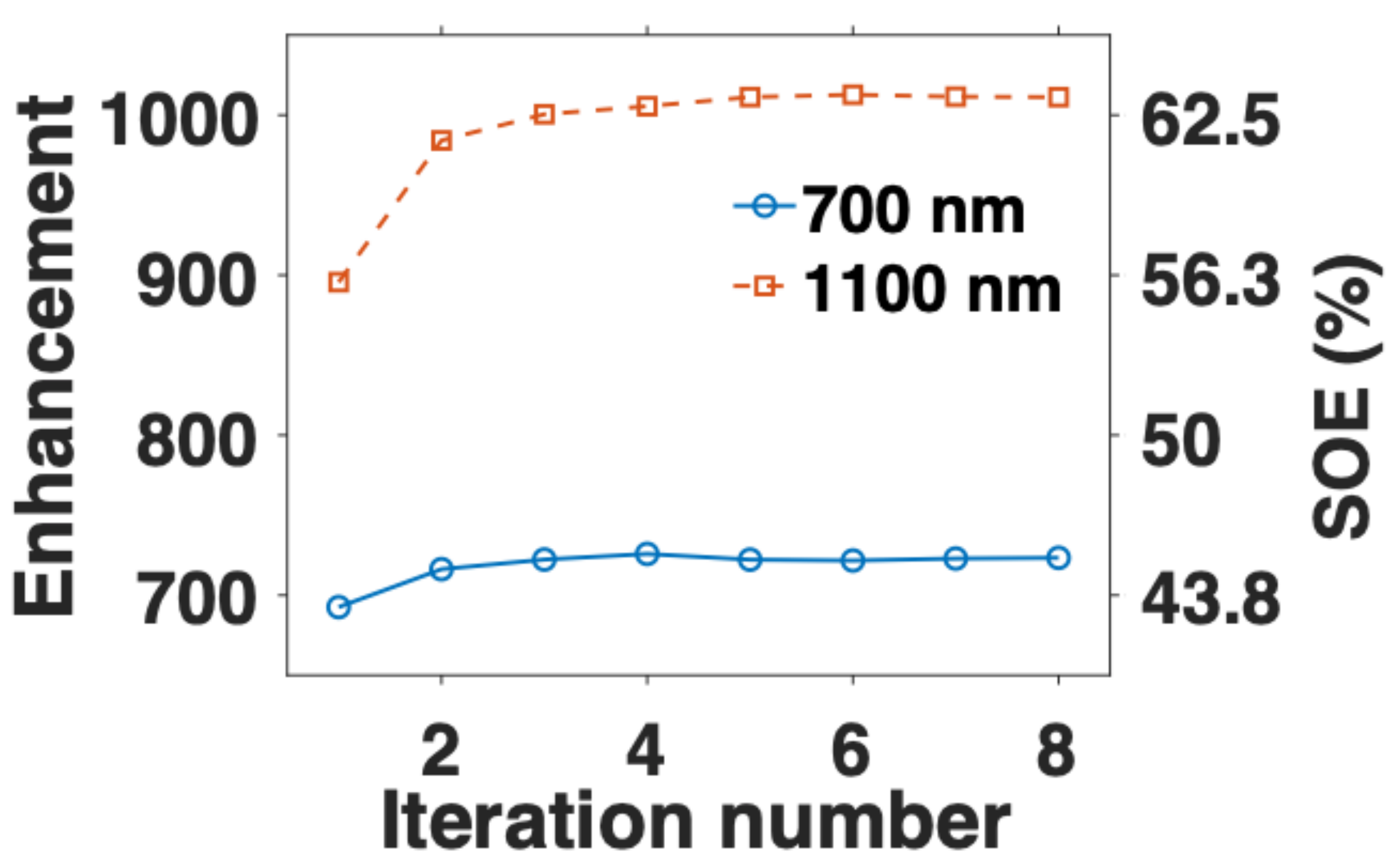}\\
      \end{tabular}
  \caption{Results of (a) OR, (b) AND, and (c) MEAN logic operations. The blue circles represent the enhancement values for 700 nm, and the red squares represent enhancement values for 1100 nm light. The dashed and solid lines are guides to the eye .}
  \label{spots}
\end{figure}

\begin{figure*}[tph]
\centering
  \includegraphics[width=170 mm]{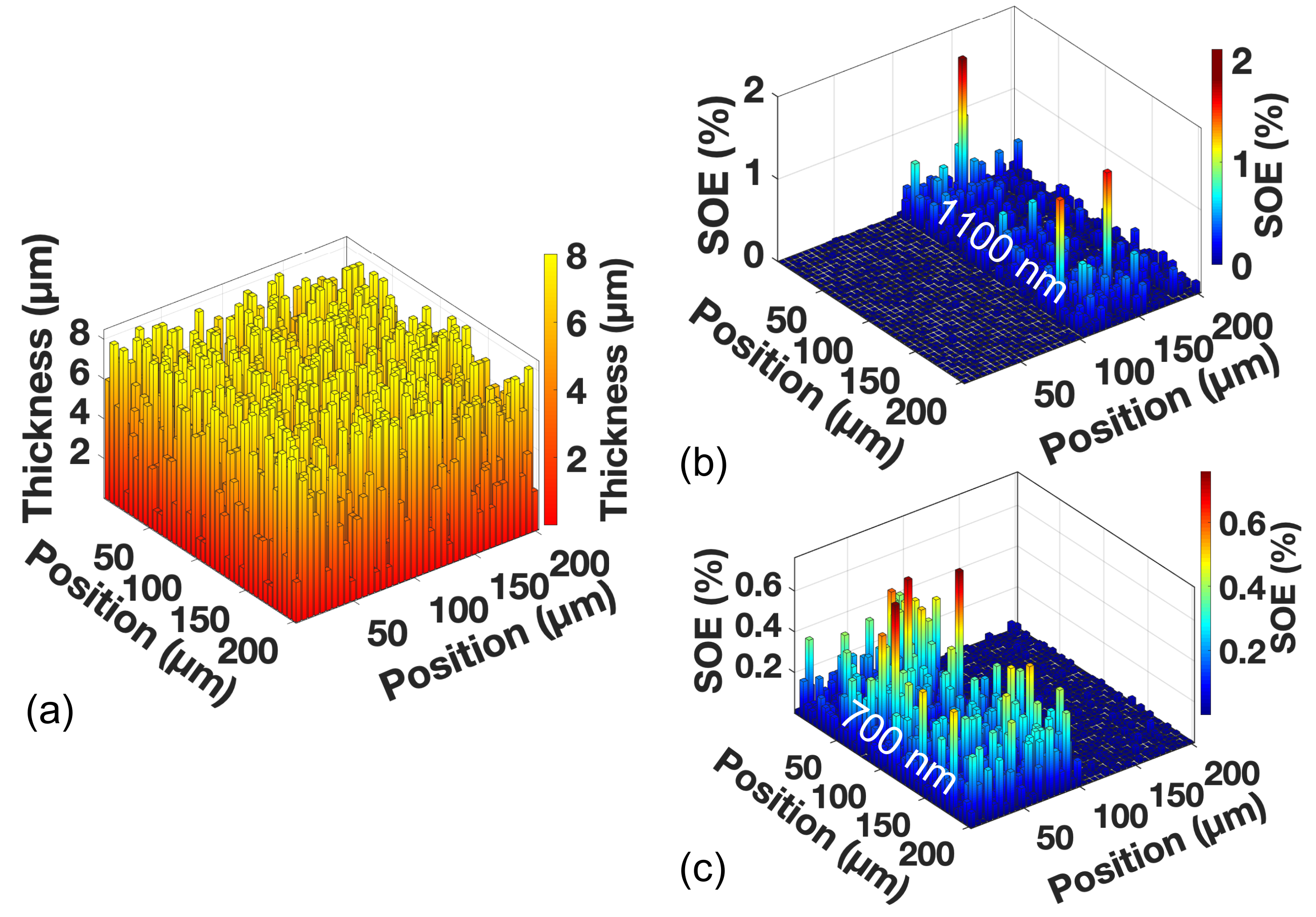}
  \caption{a) DOE thickness profile, b\&c) Distributions of SOE on the diffraction plane for two monochromatic light sources using \textit{MEAN} criterion. (b) Distribution of SOE of a light source at 1100 nm. The light source is directed on the left half of the target plane. The cumulative SOE over the target is 94\%. (c) Distribution of SOE of a light source at 700 nm. The light source is directed on the right half of the target plane. The cumulative SOE over the target plane is 92\%.}
  \label{tworegion}
\end{figure*}

We first inspect \textit{OR} logic operation to design a DOE to guide two continuous light sources with wavelengths of 700 nm and 1100 nm on two targets. Fig. \ref{spots}.a shows that result of enhancements/SOEs when the optimization is performed with \textit{OR} logic operation. The algorithm with the \textit{OR} case updates the DOE profile when SOE/enhancement of either one or both two light sources increases at each sub-iteration. Therefore, at each sub-iteration, the algorithm finds another local maximum for each wavelength. We do not observe saturation of the enhancements/SOEs after 8 iterations (takes 4.6 hours on a desktop PC), and we obtain maximum enhancement around 5 iterations. At the end of the optimization, we conclude that enhancements of 303 and 316 at wavelengths 700 nm and 1100 nm, respectively. In terms of SOE, we manage to direct 19\% and 20\% of the incident light at 700 nm and 1100 nm to a single pixel, respectively. The SOE values will increase if the target area is greater, which will be shown for the DOEs designed for broadband operation.

When DOE thickness profile is optimized with \textit{AND} criterion at each sub-iteration, SOEs/enhancements of all design wavelengths are improved as seen in Fig. \ref{spots}.b. SOEs/enhancements of both light sources saturate at 6 iterations, and we concluded that 4 hours are sufficient to calculate an optimal DOE to concentrate two light sources with \textit{AND} logic operation. At the end of the optimization, 676 and 656 enhancements (42\% and 41\% in terms of SOE) on the targets are achieved for 700 nm and 1100 nm, respectively. 

Lastly, enhancements of two light sources with \textit{MEAN} criterion are presented in Fig. \ref{spots}.c. The DOE thickness profile is updated when the average enhancement/SOE of two light sources increases at each sub-iteration. As expected, one of two light sources is focused very strongly than the other due to mean of them that increases at each sub-iteration. At the end of all the main iterations, enhancements of 723 and 1011 (45\% and 63\% in terms of SOE) are obtained for 700 nm and 1100 nm, respectively. Again SOEs/enhancements of both light sources saturate at 6 iterations, and we concluded that 4 hours are sufficient to calculate an optimal DOE to concentrate two light sources with \textit{MEAN} logic operation. It is observed that the \textit{MEAN} criterion among other logic operations gives the highest average SOE for these light sources. 

The ultimate pixel number is defined by A/$\lambda^2$, where A is the area of DOE. Here, we use 1600 pixels (40x40) for our DOEs. The SOEs and enhancements values will surely increase with more pixels \cite{Vellekoop2008}. However, this would be beyond the precision limits of fabricating DOEs for the solar cells. Moreover, we obtain better SOE for the longer wavelength. It is expected that control over longer wavelengths is harder given that a thicker medium is required for getting a similar phase shift when compared to a shorter wavelength. However, in our case, we choose \SI{1}{\micro\metre} step size in DOE thickness optimization. This resolution is a coarse step for the shorter wavelength whereas the step size is sufficiently small for the long wavelength. As a result, both the SOE and the enhancement values are greater for the longer wavelengths. Here, we chose the number of pixels, DOE dimensions, and resolution limits considering direct laser writing process which can be applied to large scale fabrication. Greater enhancement over the shorter wavelengths is definitively possible using more precise fabrication methods \cite{Tokel2017}.

Next, we perform similar calculations to guide the same light sources (700 nm and 1100 nm) on a larger target (0.1x0.2 mm). The larger target area inherently involves more degree of freedom for spectral splitting of broadband light source. For this purpose, we use \textit{AND} and \textit{MEAN} logic operations. With the \textit{MEAN} criterion, 700 nm and 1100 nm sources are focused with higher SOEs: 92\% and 94\%, respectively (Fig. \ref{tworegion}). In Fig. \ref{tworegion}.a the DOE thickness profile is presented. In sub-figures (b) and (c) of Fig. \ref{tworegion}, it is obviously seen that the distribution of light outside the target area is considerably weak. 

Optimizing DOE for broadband light is quite a challenge. Using 1 nm wavelength steps between 400 nm - 1100 nm with 40x40 pixels each of which can be obtained 8 different thickness values results in minimum 9x10$^{6}$ variables to optimize, and grows exponentially with number of optimization variables. The calculation of such optimization will take approximately 89 days on a desktop PC, which is computationally expensive for optimizing DOEs.

\begin{figure}
  \centering
     {\textbf{}}\includegraphics[width=70mm]{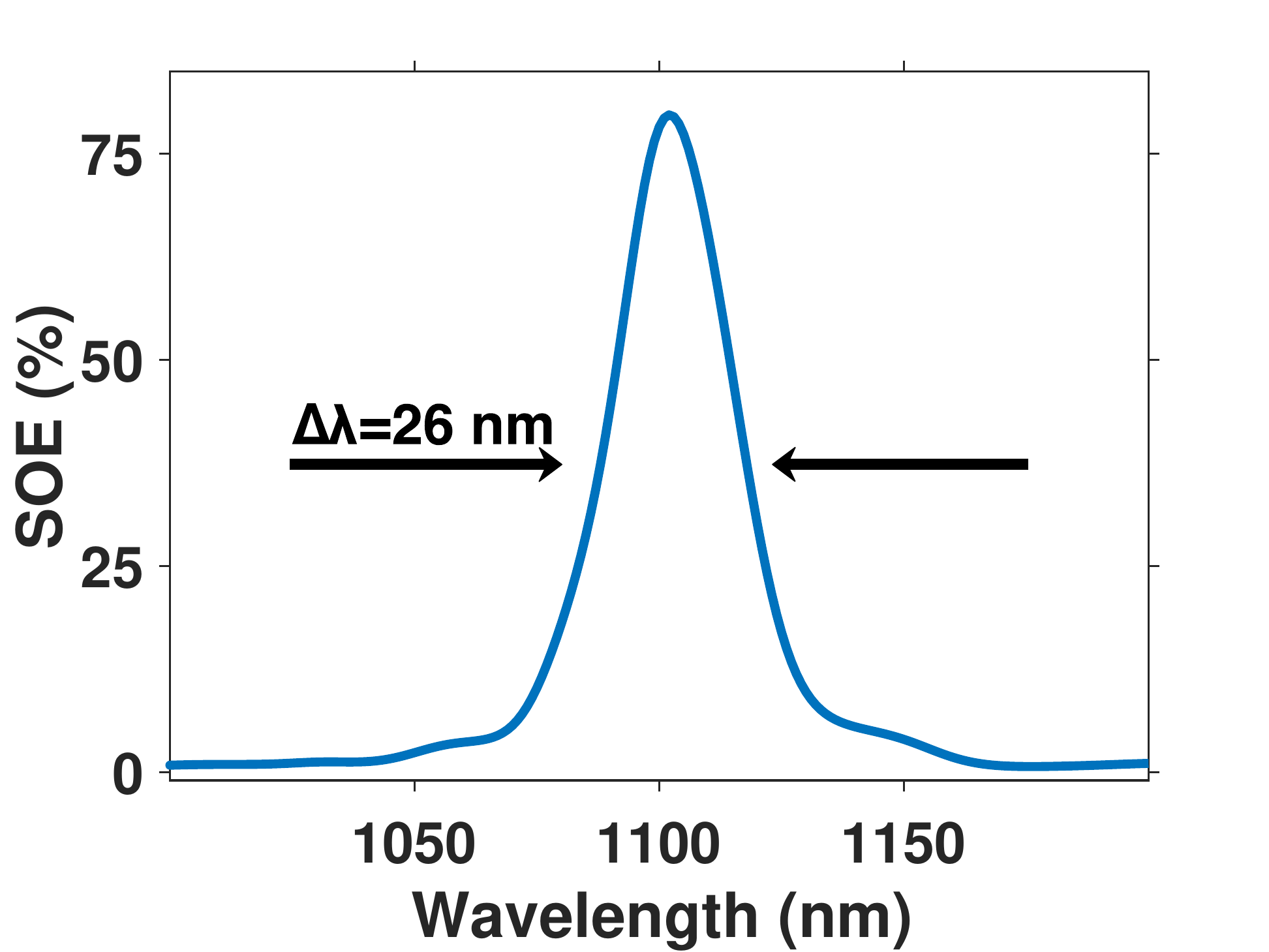}\\
  \caption{The broadband response of the DOE designed for guiding 1100 nm on a single pixel on the output plane.}
  \label{Bandwidth}
\end{figure}

Here, we introduce an effective bandwidth approach to overcome the extremely lengthy computational duration. For this purpose, we first design a DOE for 1100 nm and then calculate its response for a variety of input wavelengths. Fig. \ref{Bandwidth} shows the output spectrum of the DOE designed for 1100 nm. As can be seen in Fig. \ref{Bandwidth}, the DOE has a response bandwidth of $\Delta\lambda$=26 nm (FWHM). Using half of this bandwidth as our step in wavelength we design a DOE for spectral splitting of broadband light.

\begin{figure}
  \centering
  \begin{tabular}{cc}
     \includegraphics[width=70mm]{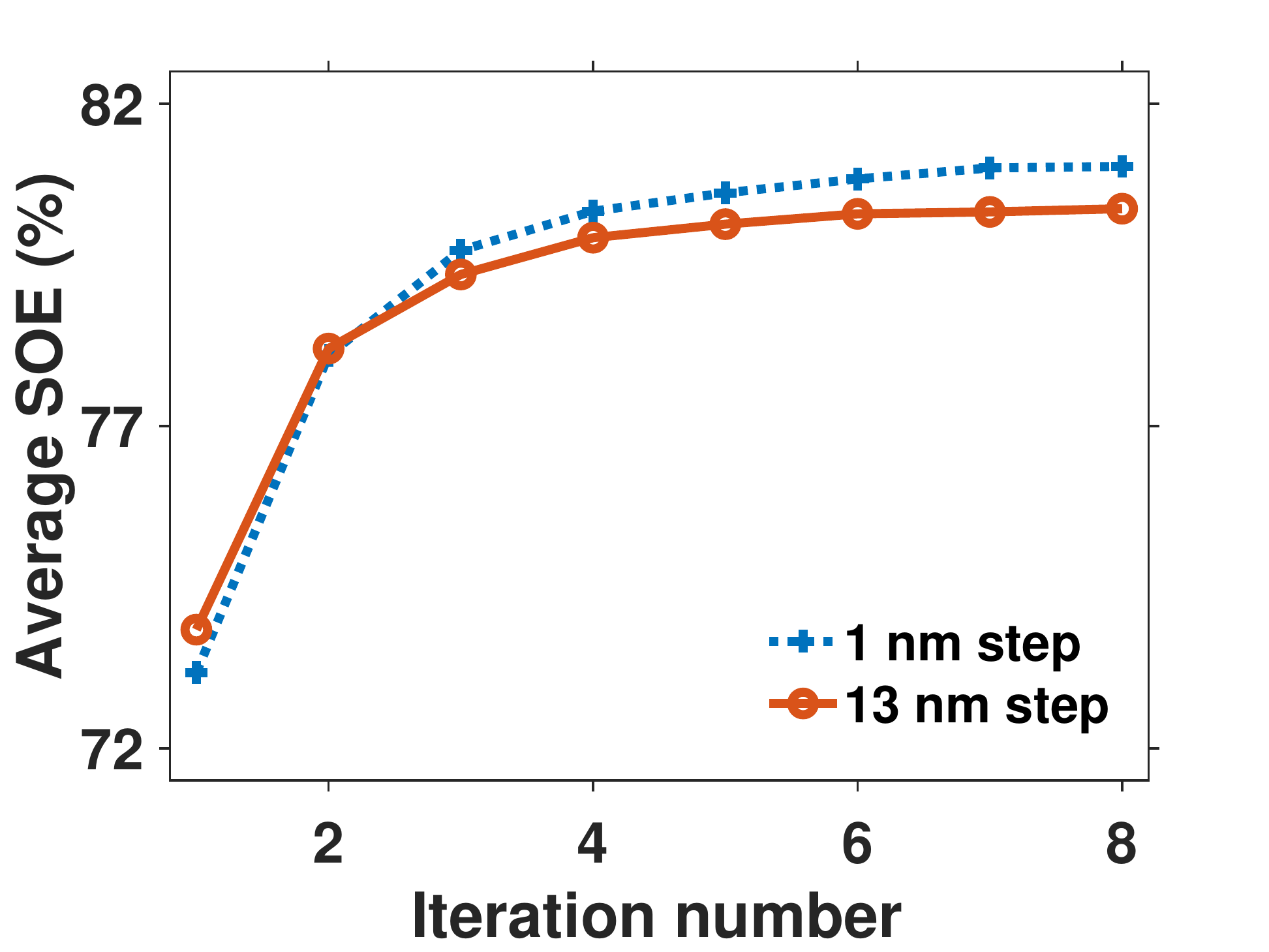}\\
  \end{tabular}
   \caption{Evolution of average SOE for band 1051 nm - 1100 nm with the bandwidth approach which is 13 nm wavelength step size (red solid lines) and with a finer wavelength step size of 1 nm (blue dotted lines). The solid and dashed lines are guides to the eye.}
  \label{Comp}
\end{figure}

\begin{figure}
  \centering
  \begin{tabular}{cc}
     \includegraphics[width=70mm]{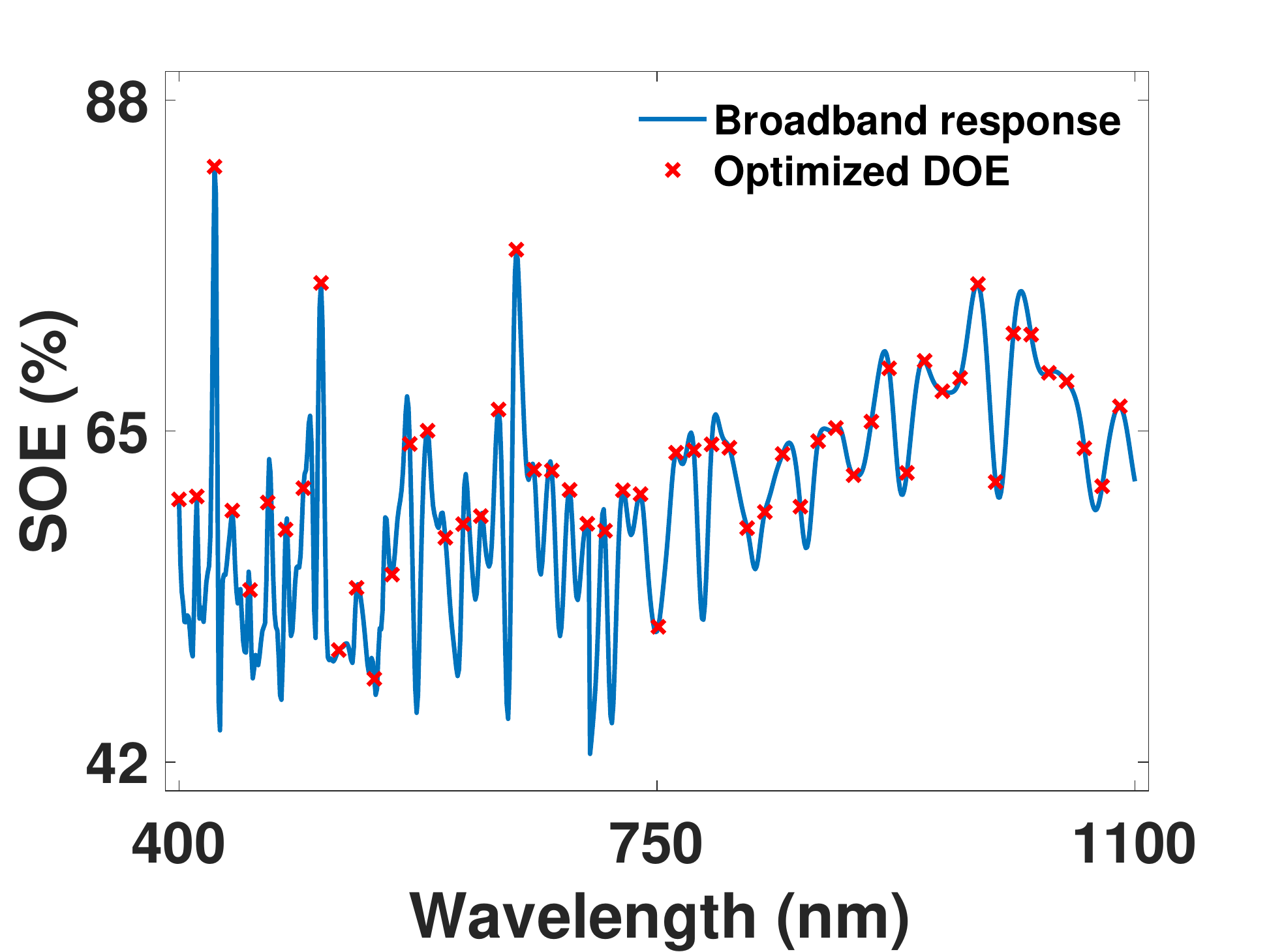}\\
  \end{tabular}
   \caption{Wavelength-selective SOE of a broadband DOE optimized with a bandwidth of 13 nm wavelength steps to disperse the broadband light. Red cross points show SOE of each design wavelength. Blue solid line represents the broadband response of the optimized DOE in 1 nm wavelength step.}
  \label{Broadband}
\end{figure}

\begin{figure}
  \centering
  \begin{tabular}{cc}
     {\textbf{(a)}}\includegraphics[width=70mm]{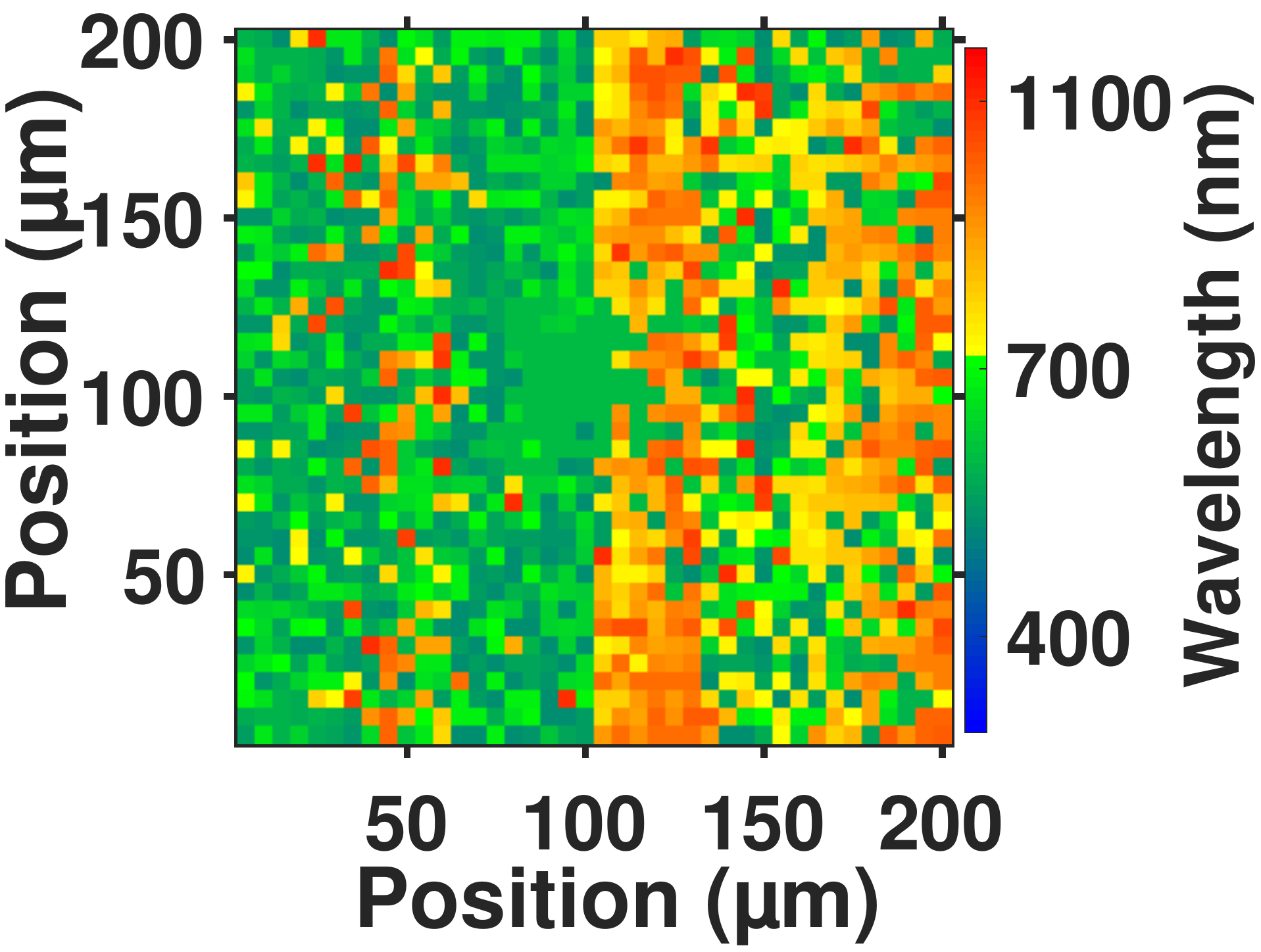}\\
     {\textbf{(b)}}\includegraphics[width=70mm]{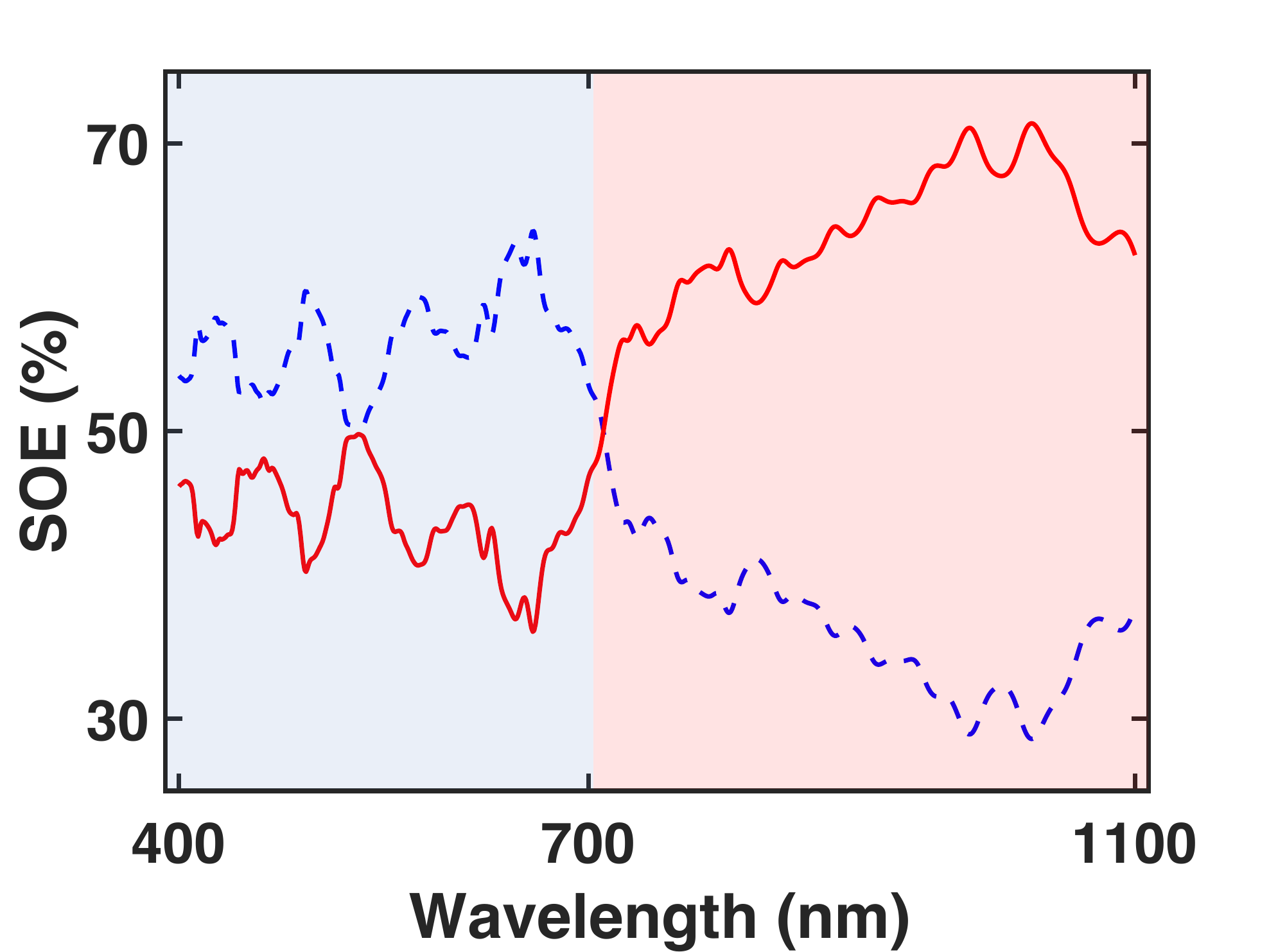}\\
  \end{tabular}
  \caption{(a) The distribution of wavelength of the broadband light source having the highest SOE over the target. Each target pixel represents the wavelength of a monochromatic light source having maximum SOE on the corresponding target, (b) SOE of each wavelength using an optimized DOE with 13 nm wavelength steps. Blue dashed and red solid lines show SOE of each light source.}
  \label{Broaddist}
\end{figure}

In order to test the effectiveness of our bandwidth approach, we concentrate input wavelengths between 1051 nm and 1100 nm. We first optimize a DOE using 1 nm wavelength step size and then we compute another DOE with the bandwidth approach: 13 nm wavelength step size. Later, the performance of both DOEs are tested under broadband illumination. The results are provided in Fig. \ref{Comp}. We observe that our bandwidth approach demonstrates similar results to the DOE optimized with 1 nm wavelength step size. The reason is that 1 nm wavelength step size introduces too many parameters to optimize and the optimization algorithm get stuck in a local maximum rather than reaching a global maximum. With bandwidth approach average SOEs in the band 1051 nm - 1100 nm show similar values at each iteration. Thus, using the effective bandwidth approach we are able to overcome the extensive computational time while retaining the performance.

Using a bandwidth of 13 nm, 54 distinct wavelengths between 400 nm - 1100 nm are controlled to direct to a specified position on the target plane. To obtain a broadband DOE with these light sources, we use the \textit{MEAN} criterion. The algorithm uses average SOE of wavelengths between 400 nm - 700 nm as well as the average SOE of wavelengths within 701 nm - 1100 nm to optimize a single DOE structure. Considering 13 nm steps in wavelength, a single broadband DOE is obtained to spectrally split the broadband light within only 8 days calculation time. Fig. \ref{Broadband} shows SOEs at each design wavelength. As a result of this study, the average SOEs of sub-bands are obtained as 56\% for the band: 400 nm - 700 nm and 62\% for the band: 701 nm - 1100 nm. In the same figure, the result of the same DOE is used to spatially separate the broadband light with 1 nm resolution is presented. Averages for the sub-bands between 400 nm - 700 nm and 701 nm -1100 nm are 56\% and 63\%, respectively. Therefore, we conclude that the DOE designed for splitting 54 design wavelengths performs similarly with 701 wavelengths when illuminated with broadband light, as can bee seen in Fig. \ref{Broadband}. 

In Fig. \ref{Broaddist}.a the distribution of wavelengths of light sources having the highest SOE on each the target pixel is demonstrated. Fig. \ref{Broaddist}.a shows that we can spectrally split broadband light using our effective bandwidth approach. The separation of colors is distinctly visible on the target plane. In Fig. \ref{Broaddist}.b quantitative value of SOE for each wavelength is presented. Blue dashed line peaks in the sub-band 400 nm - 700 nm on the blue region of the target plane and dips on the red region of the target plane. Red continuous line peaks in the sub-band 701 nm - 1100 nm on the red region of the target plane and dips on the blue region of the target plane. The average SOEs of sub-bands are obtained as 56\% for the band: 400 nm - 700 nm and 63 \%for the band: 701 nm - 1100 nm.

\section{\label{sec:level5}Conclusions\protect 
\lowercase{} }

We here put forward three logic operations for the local search optimization algorithm to design DOEs for spectrally splitting broadband light. We demonstrate that two light sources are focused successfully to target area using a single DOE via especially the \textit{MEAN} logic operation. In our study, we aim to optimize DOEs that are compatible with today's large scale microfabrication methods such as direct laser writing. By increasing the resolution in thickness of each pixel (similar to Ref. \cite{Xiao2016, Lin2016, Huang2016}) we can obtain greater SOE and enhancement values. By introducing the bandwidth approach, we designed a single broadband DOE that performs spectral splitting and concentration of broadband light. Our approach enables us to decrease computation time by 11 times while retaining the performance. Our results enable wide-spread usage of DOEs in many fields, especially in solar energy and spectroscopy. DOEs also promise to replace and outperform several optical elements such as lenses, filters, gratings, and metalenses \cite{Chen2018}. We project that many applications that require control over broadband light can benefit from DOEs that are now accessible using our bandwidth approach in a much shorter time period.

\begin{acknowledgments}
This study is financially supported by The Scientific and Technological Research Council of Turkey (TUBITAK), grant no 118F075. Ph.D. study of Alim Yolalmaz is supported by The Scientific and Technological Research Council of Turkey (TUBITAK), with grant program of 2211-A.
\end{acknowledgments}

\bibliography{MyReferences}

\end{document}